\documentclass{amsart}
\usepackage{ulem}

\def \A {\mathcal{A}}
\def \a {\alpha}
\def \f {f_b(k)}
\def \l {\lambda}
\def \S {\Sigma}
\def \Z {\mathbb{Z}}

\def \vbar {\,|\,}
\newcommand{\bb}[1]{\textbf{#1}}
\newcommand{\ub}[1]{\underline{\textbf{#1}}}
\newcommand{\uu}[1]{\underline{#1}}
\newcommand{\so}[1]{\sout{#1}}

\newtheorem {theorem}{Theorem}
\newtheorem {lemma}    [theorem]{Lemma}
\newtheorem {corollary}[theorem]{Corollary}
\theoremstyle{definition}
\newtheorem*{definition}{Definition}
\newtheorem*{example}   {Example}
\newtheorem*{remark}    {Remark}
\newtheorem*{remarks}   {Remarks}
\newtheorem*{acknowledgements}{Acknowledgements}

\begin{document}

\title{Minimal DFAs for Testing Divisibility}
\author{Boris Alexeev}
\date{September 28, 2003}
\maketitle


\section{Statement of the Problem}
\label{problem_statement}

The following exercise is typical in introductory texts on deterministic finite automata (DFAs): ``produce an automaton that recognizes the set of binary strings that, when interpreted as binary numbers, are divisible by $k$.'' For example, exercise~1.30 in \cite{sipser} asks the student to prove that the language \{$x\vbar x$ is a binary number that is a multiple of $k$\} is regular for each $k\ge1$; explicitly presenting an automaton is the easiest solution.

The traditional (and correct) answer constructs a $k$-state automaton that keeps track not only of divisibility by $k$, but also the current residue modulo $k$. For example, if the input read was $1101$, the machine would remember ``$13 \bmod k$''. The transitions between states are simple: if the automaton's current state is ``$r \bmod k$'', and the input symbol read is ``0'', it moves to state $(2r) \bmod k$; if the input symbol read is ``1'', it moves to state $(2r+1) \bmod k$.

(This example also generalizes to bases other than binary. Furthermore, even if the input string is encoded in base $b$, the canonical DFA will still have $k$ states. It will, however, contain $b$ transitions from each state.)

The traditional answer, unfortunately, in general fails to produce a \textit{minimal} DFA. This paper addresses the considerably more difficult question of ``how many states does a minimal DFA that recognizes the set of base-$b$ numbers divisible by $k$ have?'' We denote this number by $\f$ and derive a closed-form expression; in the proof, we also describe the states of the minimal DFA in more detail.

The function $\f$ may be computed by algorithmic means. The author used two implementations of the Hopcroft minimization algorithm: an original Perl program and the highly-optimized AT\&T FSM Package\texttrademark. According to experts in the field, no prior work addresses the general case of this problem except through such computational alleys.


\section{Interesting Patterns}
\label{interesting_patterns}

The function $\f$ exhibits very curious behavior. One interesting pattern considers $\f$ with $b$ fixed and $k = x \cdot y^z$ for increasing values of $z$.

\begin{example}
Table of $\f$ for $b=6$ and $k=2^z$. (That is, $x=1$, $y=2$, and $z$ ranges from 0 to 10.)
\begin{center}
\begin{tabular}{|r|r|r|r|r|r|r|r|r|r|r|r|}
\hline
                      $z$ & 0 & 1 & 2 & 3 & 4  &  5 &  6 &   7 &   8 &   9 &   10 \\ \hline
                    $2^z$ & 1 & 2 & 4 & 8 & 16 & 32 & 64 & 128 & 256 & 512 & 1024 \\ \hline
               $f_6(2^z)$ & 1 & 2 & 3 & 5 & 8  & 12 & 20 &  29 &  45 &  72 &  104 \\ \hline
$f_6(2^{z+1}) - f_6(2^z)$ & 1 & 1 & 2 & 3 & 4  &  8 &  9 &  16 &  27 &  32 &   64 \\ \hline
\end{tabular}
\end{center}
The successive differences of $f_6(2^z)$ are the powers of $2$ and $3$, sorted in increasing order!
\end{example}

\begin{example}
Table of $\f$ for $b=2^2\cdot 5=20$ and $k=30 \cdot 5^z$. (That is, $x=30$, $y=5$, and $z$ ranges from 0 to 6.)
\begin{center}
\begin{tabular}{|r|r|r|r|r|r|r|r|}
\hline
                       $z$ &  0 &   1 &   2 &    3 &     4 &     5 &      6 \\ \hline
            $30 \cdot 5^z$ & 30 & 150 & 750 & 3750 & 18750 & 93750 & 468750 \\ \hline
    $f_{20}(30 \cdot 5^z)$ &  4 &   6 &  14 &   26 &    58 &   118 &    246 \\ \hline
$f_{20}(30\cdot5^{z+1}) - f_{20}(30\cdot5^z)$
                           &  2 &   8 &  12 &   32 &    60 &   128 &    300 \\ \hline
\end{tabular}
\end{center}
Here, the successive differences of $f_{20}(30\cdot 5^z)$ come in increasing order from two sequences: $\{2 \cdot 4^m\} = \{2, 8, 32, 128, \ldots\}$ and $\{12 \cdot 5^m\} = \{12, 60, 300, \ldots\}$.
\end{example}

We observe that the function $\f$ manages to pick terms, in increasing order, from two unrelated sequences! At first, it is hard to imagine a formula that would produce such a function. Investigating this bizarre behavior was the starting point for this study.

\section{Main Result}

\begin{theorem}
  \label{main_result}
  Let $\l(x, y) = \frac{x}{\gcd(x,y)}$. Then
  \begin{eqnarray*}
    \f
    & = & \l(k,b^\infty) + \sum_{\a=0}^{\infty} {\min\left\{ \l(b^\a,k), \l(k,b^\a)-\l(k,b^{\a+1}) \right\}} \\
    & = & \min_{\A\ge0}\left\{ \l(k,b^\A) + \sum_{\a=0}^{\A-1}{\l(b^\a,k)} \right\} \\
    & = & \l(k,b^{\A_0}) + \sum_{\a=0}^{\A_0-1}{\l(b^\a,k)},
  \end{eqnarray*}
  where $\A_0$ is the smallest nonnegative integer $\a$ satisfying $\l(k,b^\a)-\l(k,b^{\a+1})<\l(b^\a,k)$.
\end{theorem}
\begin{remarks}
  The function $\l(x,y)$ is not symmetric; indeed, $\l(x,y)=\l(y,x)$ if and only if $x=y$.

  We use the notation $\l(k, b^\infty)$ to denote $\l(k, b^\a)$ for sufficiently large $\a$; similarly, the infinite sum can be truncated when $\l(k,b^\a) - \l(k,b^{\a+1}) = 0$. This equality certainly holds for $\a \ge \log_2k$.
  
  Lemma~\ref{lemma_expression_equivalence} shows that the three expressions in the theorem are equivalent.
\end{remarks}

To understand the expressions of $\f$ in the theorem, we may draw a table listing $\a$, $\l(b^\a,k)$, $\l(k,b^\a)$, and $\l(k,b^\a)-\l(k,b^{\a+1})$. The first and third expressions may be understood fairly simply as written. However, the second expression is more difficult; it states that $\f$ is the minimal sum one can obtain by summing zero of more elements of the form $\l(b^\a,k)$ (as $\a$ ranges from $0$ to $\A-1$) and then the following value of $\l(k,b^\a)$ (that is, $\a=\A$).

\begin{example}
$b=6$, $k=16=2^4$: We can calculate $\f$ with any of the expressions above (for the third, use $\A_0=2$). The minimal terms of the first expression appear underlined below; simultaneously, the minimal ``path'' $8=1+3+4$ (in terms of the second formula above) is indicated in boldface. Note that other paths such as $15=1+3+9+2$, $9=1+8$, and $16=16$ (the trivial path $\A=0$) yield non-minimal sums.
\begin{center}
\begin{tabular}{|r|r|r|r|r|r|r|r|}
\hline
            $\a$ & 0      & 1      & 2      & 3      & 4      & 5      & 6      \\ \hline
    $\l(b^\a,k)$ & \ub{1} & \ub{3} & 9      & 27     & 81     & 486    & 2916   \\ \hline
    $\l(k,b^\a)$ & 16     & 8      & \bb{4} & 2      & 1      & 1      & 1      \\ \hline
$\l(k,b^\a)-\l(k,b^{\a+1})$
                 & 8      & 4      & \uu{2} & \uu{1} & \uu{0} & \uu{0} & \uu{0} \\ \hline
$\l(k,b^\infty)$ &        &        &        &        &        &        & \uu{1} \\ \hline
\end{tabular}
\end{center}
\end{example}

\section{Corollaries to the Main Result}

\begin{corollary}
  The following are upper bounds for $\f$:
  \begin{center}
  \begin{tabular}{rll}
    $\f$ & $\le k $                       & $ = \l(k,b^0) $ \\
    $\f$ & $\le 1 + \dfrac{k}{\gcd(k,b)}$ & $ = \l(b^0,k) + \l(k,b^1) $ \\
    $\f$ & $\le 1 + \dfrac{b}{\gcd(b,k)} + \dfrac{k}{\gcd(k,b^2)} $
                                          & $ = \l(b^0,k) + \l(b^1,k) + \l(k, b^2) $
  \end{tabular}
  \end{center}
\end{corollary}
\begin{proof}
  These follow immediately from the second expression in Theorem~\ref{main_result}.
\end{proof}

\begin{corollary}
  The canonical DFA described in Section~\ref{problem_statement} is minimal if and only if $\gcd(k,b)=1$ or $k = 2$.
\end{corollary}
\begin{proof}
  The canonical DFA has $k$ states and hence we must determine when $\f=k$.

  If $\gcd(k,b)=1$ or $k=2$, the first expression of Theorem~\ref{main_result} immediately gives $\f=k$. Otherwise, we have $\frac{k}{\gcd(k,b)}<k-1,$ and by the previous corollary,
  $$ \f \le 1 + \frac{k}{\gcd(k,b)} < k. $$
\end{proof}

\begin{corollary}
  The successive differences of $f_6(2^z)$ are powers of $2$ and $3$, sorted in increasing order.
\end{corollary}
\begin{proof}
  Manipulation of the result of the theorem yields
  \begin{eqnarray*}
    f_6(2^z)
    & = & \l(2^z,6^\infty) + \sum_{\a=0}^{\infty} { \min \left\{ \l(6^\a, 2^z), \l(2^z,6^\a) - \l(2^z,6^{\a+1}) \right\}  } \\
    & = & 1 + \sum_{\a=0}^{\infty} { \min \left\{ 3^\a\cdot\lceil2^{\a-z}\rceil, \lfloor 2^{z-\a-1} \rfloor \right\}  } \\
    & = & 1 + \sum_{\a=0}^{z-1} { \min \left\{ 3^\a, 2^{z-\a-1} \right\}  }.
  \end{eqnarray*}
  It is not difficult to see that as one increments $ z \mapsto z+1 $, a new term of the form $\min \left\{ 3^\a, 2^{z-\a-1} \right\}$ is added, and the desired property holds.
\end{proof}
\begin{remark}
  A similar approach may be applied to the general case of $f_b(x\cdot y^z)$ for increasing values of $z$. In particular, we can easily prove the pattern we noticed in Section~\ref{interesting_patterns} for $f_{20}(30\cdot5^z)$.
\end{remark}

\begin{corollary}
\label{corollary_prime}
  If $b=p^n$ ($p$ not necessarily prime, but see the remark) and $k=p^m \cdot x$ with $\gcd(x,p) = 1$, then $\f = x + \lceil\frac{m}{n}\rceil$.
\end{corollary}
\begin{proof}
  We use the first expression of the theorem:
  \begin{eqnarray*}
    \f
    & = & \l( k, b^\infty ) + \sum_{\a=0}^{\infty} { \min \left\{ \l(b^\a, k), \l(k,b^\a) - \l(k,b^{\a+1}) \right\}  } \\
    & = & x + \sum_{\a=0}^{\infty} { \min \left\{ \lceil p^{n\a-m}\rceil, \lceil p^{m-n\a}\rceil\cdot x - \lceil p^{m-(n+1)\a} \rceil\cdot x \right\}  }
  \end{eqnarray*}
  As long as $n\a < m$, $\lceil p^{n\a-m}\rceil = 1$ and $\lceil p^{m-n\a}\rceil\cdot x > \lceil p^{m-(n+1)\a} \rceil\cdot x$. There are precisely $\lceil\frac{m}{n}\rceil$ such $\a$ (since $0 \le \a < \frac{m}{n})$, so we have
  \begin{eqnarray*}
    \f
    & = & x + \sum_{\a=0}^{\lceil\frac{m}{n}\rceil-1} {\left\{1\right\}} + \sum_{\a=\lceil\frac{m}{n}\rceil}^{\infty} {\left\{0\right\}} \\
    & = & x + \left\lceil\frac{m}{n}\right\rceil,
  \end{eqnarray*}
  as desired.
\end{proof}
\begin{remark}
  If $p$ is prime, and thus $b$ is a prime power, this corollary completely characterizes $\f$, as \textit{all} $k$ can be represented in the form $p^m \cdot x$ with $\gcd(x,p) = 1$.
\end{remark}

\section{Proof of the Main Result}

\begin{lemma}
  \label{lemma_expression_equivalence}
  The three expressions of Theorem~\ref{main_result} are equivalent.
\end{lemma}
\begin{proof}
  By looking at the powers of a fixed prime, we see that $\l(b^\a,k)$ and $\gcd(k,b^\a)$ are increasing (not necessarily strictly) functions of $\a$. It is also easy to show that\linebreak$\gcd(k,b^{\a+1})/\gcd(k,b^\a)$ is decreasing, which immediately implies that $\l(k,b^\a)-\l(k,b^{\a+1})$ is decreasing. Therefore, in the sum
$$ \sum_{\a=0}^{\infty} { \min \left\{ \l(b^\a, k), \l(k,b^\a) - \l(k,b^{\a+1}) \right\} }, $$
  one takes $\A_0$ elements from the first sequence $\{\l(b^\a,k)\}$ and then infinitely many from the second sequence $\{\l(k,b^\a) - \l(k,b^{\a+1})\}$. Telescoping the latter, one gets the other two expressions of the theorem. (The cut-off $\A_0$ is the smallest nonnegative integer $\a$ satisfying $\l(k,b^\a)-\l(k,b^{\a+1})<\l(b^\a,k)$.)
\end{proof}

\begin{proof}[Proof of Theorem~\ref{main_result}]
  Constructing a DFA directly, as in Section~\ref{problem_statement}, is often difficult because one must describe the transitions between states in addition to the states themselves. We will use the Myhill-Nerode Theorem and the accompanying theory of extension invariant equivalence relations to work with the states of the automaton only.
\begin{definition}
  Given a language (set of strings) $L$ over an alphabet $\S$, we define the extension invariant equivalence relation $\sim_L$ associated with $L$ as follows: strings $x$ and $y$ in $\S^*$ are equivalent ($x\sim_Ly$) if for any suffix $z\in\S^*$, $xz\in L$ if and only if $yz\in L$. (As is customary, $\S^*$ denotes the set of all finite strings over $\S$. Later, we use $\S^+=\S^*\setminus\{\epsilon\}$ to denote the set of nonempty strings over $\S$.)
\end{definition}
The Myhill-Nerode Theorem \cite[\textit{Thms} 3.9--10]{hopcroft_ullman} establishes that the minimal-state automaton accepting $L$ has, up to isomorphism, one state corresponding to each equivalence class of $\sim_L$. Therefore, the minimal-state automaton has exactly the number of states as the index of $\sim_L$. (In particular, a language $L$ is regular if and only if $\sim_L$ has finite index.) In addition, \bb{any} DFA recognizing $L$ can be altered by identifying (``gluing'') some states together to obtain the minimal-state automaton.

In this proof, we let $\S$ be the set of base-$b$ digits and $L$ the set of base-$b$ numbers divisible by $k$. In addition, since we work with only one language at a time, we may write $x\sim y$ rather than $x\sim_Ly$.

To begin, we will restate the problem equivalently in a way that will allow us to utilize modular arithmetic. Because the canonical DFA accepting $L$ has a state for each residue modulo $k$, the Myhill-Nerode Theorem implies that the minimal-state DFA will contain states that correspond to \textit{groups of residues} modulo $k$. Therefore, in the pursuing analysis, rather than considering strings of digits, we discuss residues; in a way, we are projecting $\S^*$ onto $\Z_k$ (in the natural manner). For example, $L$ now becomes very simple: instead of containing all numbers divisible by $k$, it contains the single residue $0\pmod k$. To complete the reduction, we need only bother ourselves with one further

\begin{definition}
  Let $r\in\Z_k$ be a residue modulo $k$ and $d\in\S$ a base-$b$ digit. We define the concatenation $rd$ to be the residue $b\cdot r+d\pmod k$. Similarly, if $d=d_{n-1}\cdots d_1d_0\in\S^+$ is a nonempty string of digits, let the concatenation $rd$ be what is obtained by successively concatenating individual digits:
$$rd\equiv b\cdot(b\cdot(\dots(b\cdot r+d_{n-1})\cdots)+d_1)+d_0)\equiv b^n\cdot r+\overline{d_{n-1}\cdots d_1d_0}\pmod k,$$
where $\overline{d}$ denotes $d$ interpreted as an integer. Of course, if $d=\epsilon$, the empty string, $rd=r\epsilon\equiv r$.

  Finally, extend $\sim_L$ onto $\Z_k$: residues $x,y\in\Z_k$ are equivalent if for any string $z\in\S^*$, $xz\equiv0\pmod k$ if and only if $yz\equiv0\pmod k$.
\end{definition}

Now, suppose $\A$ is a nonnegative integer. We will describe
\makeatletter\renewcommand\theequation{*}\makeatother
\begin{eqnarray}
  \label{eqn_package_sizes}
  \l(k,b^\A) + \sum_{\a=0}^{\A-1}{\l(b^\a,k)}
\end{eqnarray}
\textit{pre-equivalence classes}, each a group of residues, which will be a refinement of the equivalence classes of $\sim_L$.

The pre-equivalence classes we define naturally present themselves in \textit{packages}, a term we borrow from computer programming to indicate collections of classes. Altogether, there are $\A+1$ distinct packages, which we number $0,\ldots,\A$; in addition, we will sometimes refer to package~$\A$ as the distinctive \textit{package~etcetera}. These packages come in the sizes anticipated from (\ref{eqn_package_sizes}): if $0\le\a<\A$, package~$\a$ contains $\l(b^\a,k)$ pre-equivalence classes, while package~$\A$ contains $\l(k,b^\A)$ pre-equivalence classes.

We now define the packages. Suppose $0\le\a<\A$. Package~$\a$ will consist of those residues $r$ such that there exists a string $d$ of length $\a$ such that $rd\equiv0$ and no smaller $\a$ works; furthermore, these residues will be grouped according to their corresponding $d$'s. Mathematically, for each $0\le c<b^\a$ such that $\gcd(b^\a,k)\vbar c$, package~$\a$ contains the pre-equivalence class $\{x\vbar b^\a\cdot x+c\equiv0\}$, except those $x$ that appeared in package $\a-1$ or earlier. (Note that the equation $b^\a\cdot x+c\equiv0$ has a solution $x$ iff $\gcd(b^\a,k)\vbar c$.) Because there are precisely ${b^\a}/{\gcd(b^\a,k)}=\l(b^\a,k)$ such $c$ in the desired range, these packages have the stated sizes. Before we proceed, note that the \textit{union} of the pre-equivalence classes in packages~$0$ through $\a$ consists of \textbf{all} residues $x$ satisfying $b^\a\cdot x+c\equiv0$ with $0\le c<b^\a$, and no others.

Package~etcetera consists of the leftovers; mathematically, it is similar, but there is no restriction on $c$: for each $0\le c<k$ (only to avoid duplication modulo $k$), package~$\A$ contains the pre-equivalence class $\{x\vbar b^\A\cdot x+c\equiv0\}$, except those $x$ that have appeared previously. Once again, we have the necessary number of classes, since ${k}/{\gcd(k,b^A)}=\l(k,b^\A)$.

\begin{example}
  $b=6$, $k=16=2^4$: the pre-equivalence classes for $\A=2$. This value of $\A$ was chosen so that these groups correspond to the states in the minimal DFA. Strikeouts indicate that the given value of $x$ satisfies $b^\a\cdot x+c\equiv0$ but already appeared in a previous package.
\begin{center}
\begin{tabular}{|r|l|r|l|r|l|}
\hline
\multicolumn{2}{|c|}{Package $0$} &
\multicolumn{2}{|c|}{Package $1$} &
\multicolumn{2}{|c|}{Package $2$ (\textit{etcetera})}
\\ \hline
\hspace{12pt} $\mathbf{c}$ & $\mathbf{\{x\}}$ &
\hspace{12pt} $\mathbf{c}$ & $\mathbf{\{x\}}$ &
\hspace{12pt} $\mathbf{c}$ & $\mathbf{\{x\}}$                    \\ \hline
0 & \{ 0 \} & 0 & \{ \so{0}, 8 \} & 0  & \{ \so{0, 8}, 4, 12 \}  \\ \hline
\multicolumn{2}{|c|}{}
            & 2 & \{     5, 13 \} & 4  & \{ 3, 7, 11, 15 \}      \\ \hline
\multicolumn{2}{|c|}{}
            & 4 & \{     2, 10 \} & 8  & \{ \so{2, 10}, 6, 14 \} \\ \hline
\multicolumn{2}{|c|}{} & \multicolumn{2}{|c|}{}
                                  & 12 & \{ \so{5, 13}, 1, 9 \}  \\ \hline
\end{tabular}
\end{center}  
\end{example}

Recall once more from the statement of the theorem that $\A_0$ is the smallest nonnegative integer $\a$ satisfying $\l(k,b^\a)-\l(k,b^{\a+1})<\l(b^\a,k)$.

We make three separate claims:
\begin{enumerate}
\item for any $\A$, our pre-equivalence classes coincide with the equivalence classes of $\sim_L$ with two possible exceptions: some pre-equivalence classes may be empty and some pre-equivalence classes in package~etcetera may actually be equivalent (both of these would produce an overcount);
\item for $\A\le\A_0$, all the pre-equivalence classes are nonempty; and
\item for $\A\ge\A_0$, the classes of package~etcetera are actually inequivalent.
\end{enumerate}
It follows that for $\A=\A_0$, our pre-equivalence classes are precisely the Myhill-Nerode equivalence classes of $\sim_L$.

We begin by affirming (1): if two residues $r$ and $s$ are in the same class of package~$\a$, there exists no string $d$ of length less than $\a$ such that $rd\equiv0$ or $sd\equiv0$. In addition, $r\cdot b^\a\equiv s\cdot b^\a$, so for any string $d$ of length at least $\a$, we have $rd\equiv sd$. Therefore, $r$ and $s$ are equivalent, and the pre-equivalence classes are a refinement of those of $\sim_L$.

Moreover, if $r$ and $s$ are in different classes and at least one of $r$ and $s$ is not in package~etcetera, then $r\not\sim s$. Indeed, if $r$ and $s$ are in different packages, the result is obviously true. If $r$ and $s$ are in different classes of the same package~$\a$ with $\a<\A$, we can also conclude that $r\not\sim s$ because $r$ and $s$ satisfy $b^\a\cdot x+c\equiv0$ for different values of $c$; therefore, there exists a string $d$ (namely, the $d$ such that $\overline{d}=c$) of length $\a$ such that $rd\equiv0$ but $sd\not\equiv0$.

Before continuing, we note the significance of $\A_0$. If $\a\le\A_0$, then
$$ \l(k,b^{\a-1})-\l(k,b^\a)\ge\l(b^{\a-1},k)\quad\iff\quad k\cdot\frac{\gcd(k,b^{\a-1})}{\gcd(k,b^\a)} \le k - b^{\a-1}, $$
and if $\a>\A_0$, then
$$ \l(k,b^{\a-1})-\l(k,b^\a) < \l(b^{\a-1},k)\quad\iff\quad k\cdot\frac{\gcd(k,b^{\a-1})}{\gcd(k,b^\a)}  >  k - b^{\a-1}. $$

Equipped, we proceed in order to (2). Suppose $\A\le\A_0$; then, we claim that for any fixed $0<\a\le\A$ and $c$ such that $\gcd(k,b^\a)\vbar c$, there exists an $x$ satisfying
\makeatletter\renewcommand\theequation{\dag}\makeatother
\begin{eqnarray}
  \label{eqn1}
  b^\a\cdot x+c\equiv0
\end{eqnarray}
which does not satisfy $b^{\a-1}\cdot x+c'\equiv0$ with $0\le c'<b^{\a-1}$. Indeed, consider all $x$ satisfying (\ref{eqn1}) and note that these $x$ are spaced apart equally with $\frac{k}{\gcd(k,b^\a)}$ separation between consecutive solutions. Multiplying these $x$ by $b^{\a-1}$ yields (possibly duplicate) residues $b^{\a-1}\cdot x$ spaced $k\cdot\frac{\gcd(k,b^{\a-1})}{\gcd(k,b^\a)}$ apart. But, because $\a\le\A\le\A_0$,
$$ k\cdot\frac{\gcd(k,b^{\a-1})}{\gcd(k,b^\a)} \le k - b^{\a-1}, $$
whence there exists an $x$ satisfying (\ref{eqn1}) such that $(b^{\a-1}\cdot x)\mod k$ is in between $1$ and $k - b^{\a-1}$, and such an $x$ cannot satisfy $b^{\a-1}\cdot x+c'\equiv0$ with $0\le c'<b^{\a-1}$. Therefore, all of the classes of packages~$0$ through $\A$ are nonempty.

We finish with (3). Suppose $\A\ge\A_0$; it suffices to show that if $r\sim s$ and $\a$ is the minimal $\a$ such that $b^\a\cdot r\equiv b^\a\cdot s$, then $\a\le A$. Assume the contrary: $\a>\A$. Then, $r$ and $s$ are both solutions of (\ref{eqn1}) for a fixed $c$. To derive a contradiction, we again focus on the spacing of solutions of (\ref{eqn1}). So, consider all $x$ satisfying (\ref{eqn1}); they are spaced $\frac{k}{\gcd(k,b^\a)}$ apart. As before, the residues $b^{\a-1}\cdot x$ for $x$ satisfying (\ref{eqn1}) are spaced $k\cdot\frac{\gcd(k,b^{\a-1})}{\gcd(k,b^\a)}$ apart. However, because $\a>\A\ge\A_0$,
$$ k\cdot\frac{\gcd(k,b^{\a-1})}{\gcd(k,b^\a)} > k - b^{\a-1} $$
and thus there is not enough room for two distinct $(b^{\a-1}\cdot x)\mod k$ in between $1$ and $k-b^{\a-1}$. Therefore, either $b^{\a-1}\cdot r\equiv b^{\a-1}\cdot s$ or one of $r$ and $s$ satisfies $b^{\a-1}\cdot x+c'\equiv0$ with $0\le c'<b^{\a-1}$. The former contradicts the minimality of $\a$, and the second is impossible as well: without loss of generality, $r$ satisfies such an equation. But then, there exists a string $d$ of length $\a-1$ such that $rd\equiv0$. Because $r\sim s$, it follows that $rd\equiv sd\equiv0$ for a string of length $\a-1$, once again contradicting the minimality of $\a$! We have reached a contradiction in all cases, therefore our assumption was false and $\a\le\A$. Therefore, any two residues are ``distinguished'' at or before $\a=\A$, and it follows that any $r$ and $s$ in the package~etcetera are equivalent if and only if they are in the same pre-equivalence class.

At last, we are done.
\end{proof}


\begin{acknowledgements}
  The author would like to thank Professors Jason Eisner of Johns Hopkins University, Dana Scott and Klaus Sutner of Carnegie Mellon University, and Michael Sipser of MIT for answering queries about the problem and related issues and Professor Rodney Canfield of the University of Georgia for many useful conversations and attention to this work.

  For large-scale computations, when speed was crucial, AT\&T Research's FSM Package was used to compute $\f$, to complement the author's own programs.
\end{acknowledgements}

\end{document}